\documentclass{DISproc}

\def\cascade{{\sc Cascade}}
\def\pythia{{\sc Pythia}}

\begin{document}

\title{\hspace*{1.6cm}  Matrix Elements with Vetoes in the 
 \\ 
\hspace*{1.1cm}  C{\large ASCADE}  Monte Carlo Event Generator}

\author{\hspace*{3.2cm}{\slshape M.~Deak$^{1,2}$,  F.~Hautmann$^3$, H.~Jung$^{4,5}$ and 
K.~Kutak$^6$}\\[1ex]
\hspace*{1.7cm}$^1$IFT-UAM/CSIC,  
Universidad Aut{\' o}noma de Madrid, E-28049 Madrid  \\
\hspace*{1.2cm}$^2$Universidade de Santiago de Compostela,   
E-15782 Santiago de Compostela\\
\hspace*{2.5cm}$^3$Theoretical Physics, 
University of Oxford,    Oxford OX1 3NP \\
\hspace*{3.0cm}$^4$Deutsches Elektronen Synchrotron, D-22603 Hamburg\\
\hspace*{3.4cm}$^5$CERN, Physics Department, CH-1211 Geneva 23\\
\hspace*{1.6cm}$^6$Instytut Fizyki Jadrowej im H. Niewodniczanskiego,  
 PL 31-342 Krakow
 }

\contribID{xy}

\doi  

\maketitle

\begin{abstract}
  We illustrate    a study  based on a veto technique  to 
 match     parton showers and matrix elements  in the  
   \cascade\  Monte Carlo event generator,  
    and  present   a numerical application to 
  gluon  matrix elements for jet production.\\   
\vskip 0.2 cm 
\hskip 1.1 cm 
{\em Contributed at the Workshop DIS2012, University of Bonn, March 2012}   
\end{abstract}

\vskip 0.3 cm

Baseline studies of   final states containing multiple  jets 
   at the Large Hadron Collider   use       Monte Carlo 
event  generators     ---  see e.g.~\cite{hoeche11} for a recent review ---      based 
on collinear evolution of  parton   showers combined with 
  hard matrix elements.   These are 
 either   high-multiplicity   tree-level matrix elements~\cite{me},    or  
next-to-leading-order matrix elements~\cite{ma} including virtual 
emission processes,  or      
possibly,       in  the  future,    a combination of both~\cite{bauergen,lave08}. 
The parton showers take into account  collinear  
small-angle  QCD radiation,    while the 
matrix elements take into account hard large-angle radiation. 

When  the  longitudinal  momentum fractions  involved in the  production  of jets 
 become small, however,  new effects  on  jet final states   arise  from  
 noncollinear 
    corrections to  parton branching  processes~\cite{hj_rec},   due to soft  but  
     finite-angle     multi-gluon emission.  
    An example of  this   occurs   at the LHC    when 
 jets are produced at   increasingly  high  rapidities~\cite{jhep09}. 
    In order to take these corrections into account one needs~\cite{mw92}  
     transverse-momentum 
dependent  showering algorithms    coupled~\cite{hef}  to hard matrix 
elements at  fixed transverse momentum.  
 
 The \cascade\   
 Monte Carlo  event generator~\cite{cascadedocu}
   provides an implementation of this framework.  
Applications  of this  to  hard  production   in the  LHC forward region~\cite{ajaltouni}
have been  investigated in~\cite{preprint},  where   studies of   
forward-central  jet  correlations     have been  proposed. 
First LHC  measurements  of jets  at  wide rapidity separations  
 have  appeared in~\cite{cms1202,dijetratios}.   
 The approach of  \cascade\   is based on a small-$x$ expansion, 
so that 
 in order to apply it to   the highest jet p$_\perp$  it is  relevant to match it 
with  perturbative fixed-order  terms. In this article  we describe a  
study      based on    a vetoing  procedure    to combine   
 shower and matrix element contributions to jet production.  
The technique  discussed  is one of the  elements  needed  to improve  the accuracy of 
\cascade\   at high transverse momenta.

To illustrate this, 
we focus on the partonic   $ q {\bar q} $ production  process 
in the $ g g^*$ channel. This can occur by direct production from gluon-gluon 
annihilation or by decay $ g \to q {\bar q} $ following    elastic  gluon scattering. 
 When the quarks have small relative transverse  
 momentum    the two mechanisms are  effectively  of the same order in the 
 strong coupling.  
  The question of  properly  simulating these processes  
   also  arises in the case of  collinear shower Monte Carlo; but   the case of the 
 transverse momentum dependent shower  involves an additional (semi)hard  scale 
set  by the  off-shellness  of the incoming parton.  Different behaviors may be 
expected depending on the ratio $| k^2 | /  \mu^2 $, where 
$| k^2 |$ is the off-shellness and  $\mu^2 $ is the merging scale used for combining 
the different production processes. 
An approach to treat this is based on the subtractive method~\cite{jccfh01}  
  (see~\cite{jcczu} and~\cite{fhfeb07} for further applications of the method).  An analysis 
   along   these lines is reported in~\cite{michal-mor12}.  In this article we describe the 
   result of another type of calculation~\cite{preprint}, based on 
   introducing a veto  on  $ g \to q {\bar q} $  splitting above a given transverse 
   momentum scale $\mu = $p$_\perp^{(V)}$.   In this  calculation   the   gluonic 
   matrix element is combined with the vetoed branching, and added to the   hard 
   production contribution.

\begin{figure}[htb]
\vspace{92mm}
\includegraphics{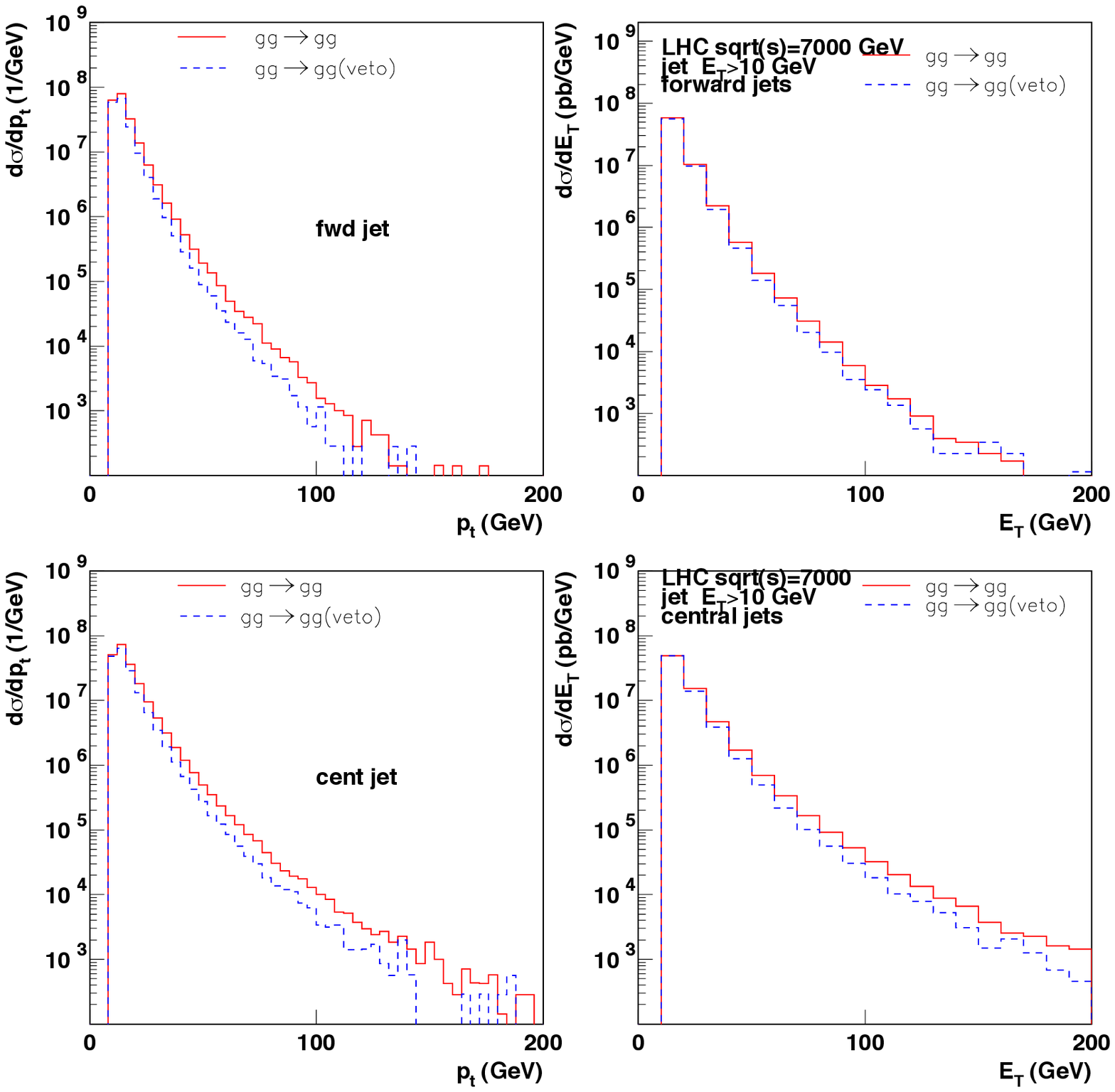}
\caption{The effect of the veto at  forward (top) and central (bottom) rapidities: 
(left) parton-level; (right) jet-level. } 
\label{fig:veto-gg}  
\end{figure}

\begin{figure}[htb]
\vspace{92mm}
\includegraphics{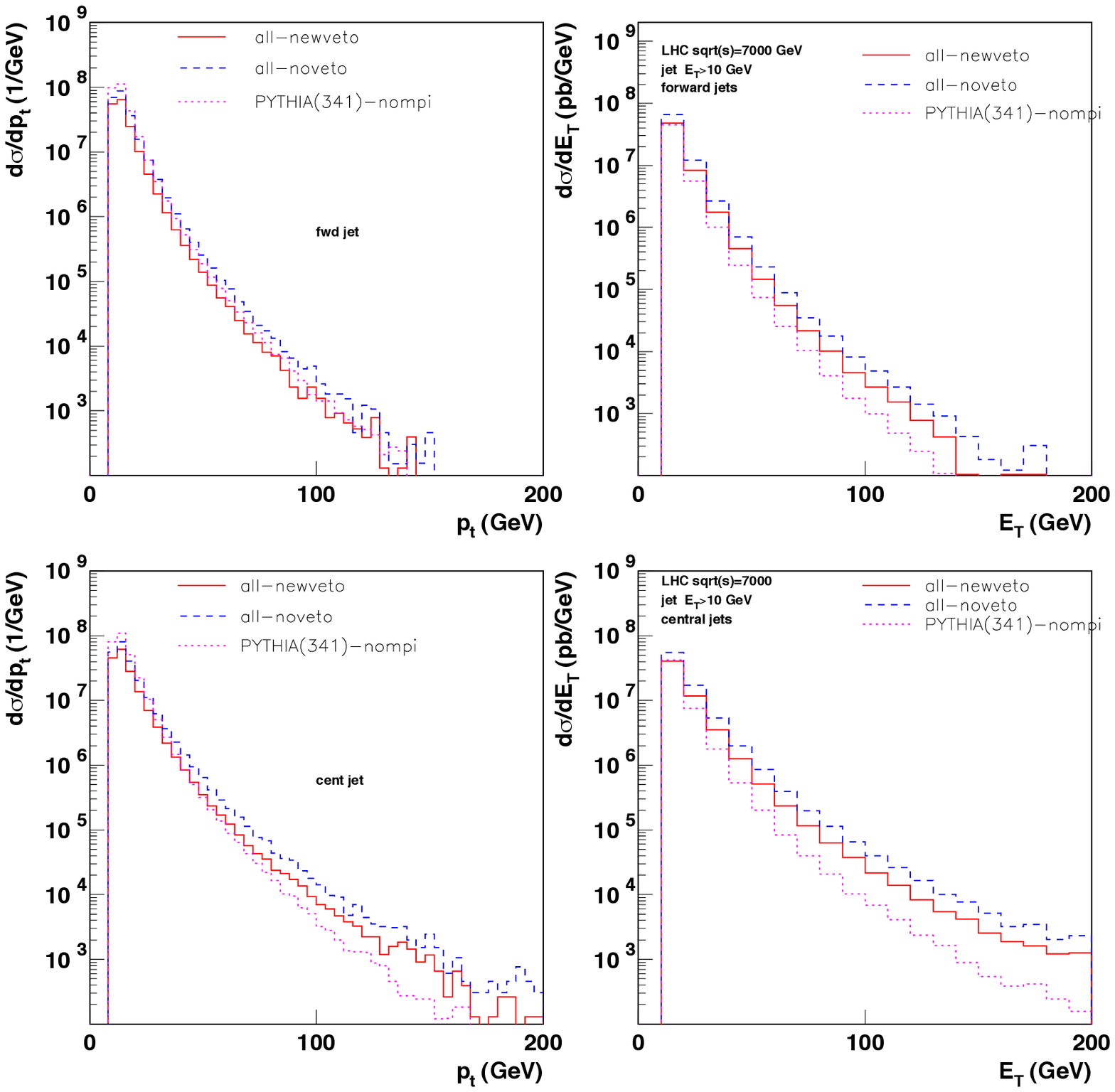}
\caption{Comparison of transverse momentum 
spectra at    forward (top) and central (bottom) rapidities: 
(left) parton-level; (right) jet-level. } 
\label{fig:veto-allch}  
\end{figure}

In Fig.~\ref{fig:veto-gg} we consider  the kinematic region~\cite{preprint}  
for  production of jets at forward and central 
rapidities, and we examine 
numerically the   effect of  the veto on the 
 gluon scattering contribution 
 both at  the level of   final state partons 
and at the level of reconstructed jets.  We see that 
in both cases  
the shape of  transverse spectra is changed by the veto. 
In Fig.~\ref{fig:veto-allch}  we 
include  all partonic  channels,  in the  same kinematic region, 
 combining   the previous contribution with  
hard production. Then the  
shape of the transverse  distribution  is  not    changed  
much as an effect of the veto, while this results into  a change in 
 normalization. 
For reference we also include the result from the \pythia\ Monte Carlo 
generator~\cite{pz_perugia} used in the LHC tune Z1~\cite{rickstune}.

In summary,   the    study  discussed in this contribution  
 introduces   vetoed decays 
   coupled  with finite-k$_\perp$ matrix elements  as 
an  approach to matching  in the case of transverse momentum dependent 
parton showers. This  is one of the ingredients 
 to extend  results of the \cascade\ Monte Carlo 
generator toward  higher p$_\perp$ jets. 
Other physical effects   will  also be    important  in 
this 
 region.    One is 
the behavior of 
the    gluon distribution for   large  $x$  at transverse momentum dependent level. 
At present   this is not very well constrained 
 in fits to  experimental data~\cite{hannes-dis12}.  
 Another  is 
  the inclusion 
of  subleading  quark contributions~\cite{unint09} to  the evolution 
of  the small-$x$ parton shower.  
In the intermediate to low p$_\perp$ range, 
studies of  the associated  mini-jet 
energy  flow~\cite{preprint,eflow-proc}  as a function of 
rapidity and  azimuthal distance  
will   be helpful to  investigate  showering 
 and  possibly  gluon rescattering~\cite{rescatt} effects. 
 We expect this  to be relevant  especially  to  analyze 
  multiple parton  interactions~\cite{bartal} and their 
 role in multi-jet production at the LHC.

\vskip 0.5 cm 

\noindent 
{\bf Acknowledgments}.   We  thank  the conveners    for the invitation  and 
excellent organization of the  meeting.


{\raggedright
\begin{footnotesize}



\end{footnotesize}
}


\end{document}